\documentclass[twocolumn,prb,notitlepage,floats,superscriptaddress,amsmath,amssymb]{revtex4-2}

\usepackage[version=3]{mhchem} 
\usepackage{bm}
\usepackage[utf8]{inputenc}
\usepackage[T1]{fontenc}
\usepackage{graphicx}
\usepackage{upgreek}
\usepackage{color}
\usepackage{ulem}
\usepackage{hyperref} 
\hypersetup{colorlinks,citecolor=blue, filecolor=blue ,linkcolor=blue , urlcolor=blue, pdftex}
\usepackage{sidecap}
\usepackage{amssymb}
\usepackage[caption=false]{subfig}

\def \FUW{Institute of Experimental Physics, Faculty of Physics, University of Warsaw, 02-093 Warsaw, Poland}
\def \Watanabe{Research Center for Electronic and Optical Materials, National Institute for Materials Science, 1-1 Namiki, Tsukuba 305-0044, Japan}
\def \Taniguchi{Research Center for Materials Nanoarchitectonics, National Institute for Materials Science,  1-1 Namiki, Tsukuba 305-0044, Japan}
\def \SingaporeEng{Department of Materials Science and Engineering, National University of Singapore, 117575, Singapore} 
\def \SingaporeInt{Institute for Functional Intelligent Materials, National University of Singapore, 117544, Singapore} 
\def \China{Guangxi Key Laboratory of Information Materials, Guilin University of Electronic Technology, Guilin 541004, China}

\begin{document}

\title{Multidimensional sensing of proximity magnetic fields via intrinsic activation of dark excitons in WSe$_2$/CrCl$_3$ heterostructure}

\author{\L{}ucja Kipczak}
\thanks{These authors contributed equally to this work.}
\email{lucja.kipczak@fuw.edu.pl}
\affiliation{\FUW}
\author{Zhaolong Chen}
\thanks{These authors contributed equally to this work.}
\affiliation{\SingaporeEng}
\affiliation{\SingaporeInt}
\author{Pengru Huang}
\affiliation{\SingaporeEng}
\affiliation{\SingaporeInt}
\affiliation{\China}
\author{Kristina Vaklinova}
\affiliation{\SingaporeEng}
\affiliation{\SingaporeInt}
\author{Kenji~Watanabe}
\affiliation{\Watanabe}
\author{Takashi Taniguchi}
\affiliation{\Taniguchi}
\author{Adam Babi\'nski}
\affiliation{\FUW}
\author{Maciej Koperski}
\email{msemaci@nus.edu.sg }
\affiliation{\SingaporeEng}
\affiliation{\SingaporeInt}
\author{Maciej R. Molas}
\email{maciej.molas@fuw.edu.pl}
\affiliation{\FUW}

\begin{abstract}
Quantum phenomena at interfaces create functionalities at the level of materials. Ferromagnetism in van der Waals systems with diverse arrangements of spins opened a pathway for utilizing proximity magnetic fields to activate properties of materials which would otherwise require external stimuli. 
Herewith, we realize this notion via creating heterostructures comprising bulk CrCl$_3$ ferromagnet with in-plane easy-axis magnetization and monolayer WSe$_2$ semiconductor.
We demonstrate that the in-plane component of the proximity field activates the dark excitons within WSe$_2$. 
Zero-external-field emission from the dark states allowed us to establish the in-plane and out-of-plane components of the proximity field via inspection of the emission intensity and Zeeman effect, yielding canted orientations at the degree range of $10^{\circ}$ $-$ $30^{\circ}$ at different locations of the heterostructures, attributed to the features of interfacial topography. 
Our findings are relevant for the development of spintronics and valleytronics with long-lived dark states in technological timescales and sensing applications of local magnetic fields realized simultaneously in multiple dimensions.
\end{abstract}

\maketitle

\section{Introduction \label{sec:Intro}}
The technology of van der Waals (vdW) assembly instigated a shift from controlling the material properties via external methods towards intra-heterostructure functionalities. 
Interfacial effects play an important role in material design as the proximity of materials with vastly different structural, chemical, optoelectronic, or magnetic properties activates novel phenomena and provides control knobs for fundamental excitations. 
Notable examples include Hall magnetometry in a metal that detects the magnetic state of an adjacent layer\cite{Hall_mag}, charge tunneling in vertical heterojunctions\cite{Proximity_tunelling}, or electronic and/or excitonic spin polarization induced by the spin splitting in the presence of a proximity magnetic field ~\cite{Hauser1969, Behera2019, Voroshnin2022, Choi2022}. 
In the domain of two-dimensional materials, it was demonstrated that interfacing a monolayer (ML) of semiconducting transition metal dichalcogenide, $e.g.$ WSe$_2$, with a layered ferromagnetic material, $e.g.$ CrBr$_3$, can drastically modify the optoelectronic properties of the semiconducting ML~\cite{Zhang2016Adv, Ghazaryan2018, Behera2019, Heissenbuttel2021}. 
Although the number of materials that could strongly interact with each other at the interface is immense, we only know a sparse representation of assembly rules leading to functionalities created via such a methodology.

Herewith, we demonstrate that the proximity between a CrCl$_3$~\cite{Hansen1958, Cable1961, Gilbertini2019} ferromagnet with an in-plane easy-axis magnetization and a non-magnetic WSe$_2$ ML semiconductor leads to a brightening of dark excitons without the application of an external planar magnetic field. 
The brightening of spin-dark excitons requires mechanisms related to the mixing of opposite spin states, which hitherto required external stimuli or intrinsic disorder. 
Our approach is different and is based on engineering functionalities on-demand via quantum materials design. 
To that end, we created heterostructures composed of a WSe$_2$ ML covered with a bulk (10 - 20~nm) ferromagnetic CrCl$_3$ layer. 
The transparency of CrCl$_3$ crystal in the energy range of the excitonic resonances in WSe$_2$ ML, combined with the encapsulation with hexagonal boron nitride (hBN), enabled inspection of the dark exciton state via narrow resonances in the photoluminesence (PL) spectra. 
The quenching of the multiexcitonic emission in WSe$_2$ ML~\cite{Courtade2017, Li2018, Chen2018, Barbone2018, Paur2019, Liu2019, Li2019, Li2019replica, Li2019momentum, Molas2019, LiuGate, LiuValley, Liu2020, He2020, Robert2021, Robert2021PRL}, driven by a charge transfer process enabled by the relative band alignment within WSe$_2$/CrCl$_3$ heterostructure, led to a formation of PL spectra dominated by the dark exciton resonance. 

\begin{figure*}[!th]
		\subfloat{}%
		\centering
		\includegraphics[width=1 \linewidth]{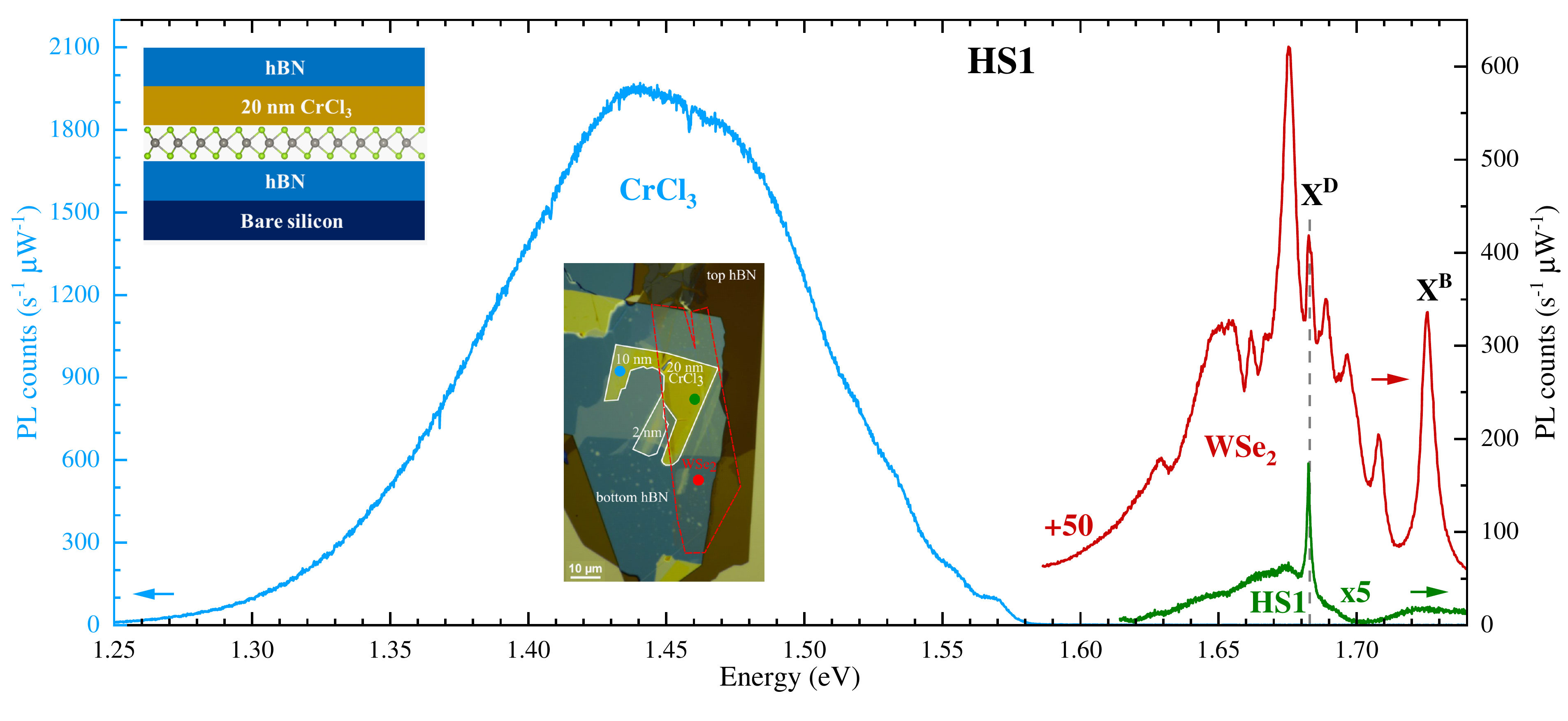}
    	\caption{Photoluminescence spectra of three structures: hBN/CrCl$_3$/hBN with a 10 nm thick CrCl$_3$ film (blue), a hBN/WSe$_2$/hBN with WSe$_2$ monolayer (red) and heterostructure hBN/WSe$_2$/CrCl$_3$/hBN (green) measured on the sample referred to as HS1 at low temperature ($\textrm{T}=5~\textrm{K}$), using excitation energy 2.41~eV and laser power of 15~$\upmu$W. 
        The spectra are vertically shifted or multiplied by scalling factor for clarity.
        The inset in the top-left corner shows the schematic side view of the sample with the indication of consecutive layers. 
        The inset in the central part of the figure demonstrates the optical image of the heterostucture. 
        The outlines for individual flakes are added as a guide to the eye. 
        Colored dots in the optical image indicate spatial positions corresponding to the photoluminescence spectra of different material configurations.}
		\label{fig:1}
\end{figure*}

The new character of the PL response transforms the WSe$_2$ ML into a sensor of magnetic fields. 
The relative contribution of the dark exciton state to the PL emission allowed us to estimate the in-plane component of the magnetic field, while the fine-structure splitting, modified by the Zeeman effect, provided an accurate value of the out-of-plane component. 
Our analysis demonstrated that at the locations of strongest enhancement of the dark exciton states, there existed a canted magnetization in the degree range of $10^{\circ}-30 ^{\circ}$ with respect to the WSe$_2$ ML, which we attribute to the three-dimensional topography structures naturally appearing at van der Waals interfaces.

The inspection of the optoelectronic properties of hBN/WSe$_2$/CrCl$_3$/hBN heterostructures demonstrates that the technological advantages of dark exctions, such as their long lifetimes and spin coherence, could be accessed intrinsically via the material design. 
This was possible through a synergistic design of vdW interfaces, combining proximity fields from a ferromagnet, charge transfer through the relative band alignment, and suppression of dielectric inhomogeneities provided by an atomically flat insulator.

\section{Results \label{sec:Experimnet}}

We inspected the optical response of hBN/WSe$_2$/CrCl$_3$/hBN heterostructures (HS) created via mechanical exfoliation and stacking of consecutive layers (see the Methods section for detailed information on the sample fabrication process). 
The hBN encapsulation enables inspection of spectrally narrow resonances that form a family of excitonic species within the semiconducting WSe$_2$ ML characterized by various charge state, spin/subband contribution, and/or phonon-replicated resonances \cite{Liu2020, He2020}. 
The detailed characteristics of the excitonic response are unveiled at low temperature.
 Notably, the overlap of the consecutive layers in HS creates three types of material stacking configurations: hBN/CrCl$_3$/hBN, hBN/WSe$_2$/hBN, and hBN/WSe$_2$/CrCl$_3$/hBN. 
The corresponding PL spectra at $T$=~5~K are presented in Fig.~\ref{fig:1}.
This allows us to study the optical response of individual materials and the HS in a comparative manner. 

The PL spectrum of the 10-nm-thick CrCl$_3$ flake shows a broad-band optical response distributed from about 1.25~eV to almost 1.60~eV, showing spectral characteristics akin to those reported previously~\cite{Cai2019}. 
This type of emission is typical for Cr-based trihalides of the chemical formula CrX$_3$ (X~=~Cl,~Br,~I) due to the emergence of Frenkel type of excitons~\cite{Seyler2018_I3, Acharya2022}. 
CrX$_3$ emission spectra can be interpreted in the regime of large Huang-Rhys factors when considering a one-dimensional Franck-Condon model~\cite{Magda@2023}. 
This means that the oscillator strength for the zero-phonon line (indicative of the electronic excitation within the molecular-like state in the crystal) is negligible, while the series of its phonon replicas form a broad emission band.
The PL spectrum for encapsulated WSe$_2$ ML displays several emission lines that have been attributed to recombination pathways of different excitonic species~\cite{Courtade2017, Li2018, Chen2018, Barbone2018, Paur2019, Liu2019, Li2019, Li2019replica, Li2019momentum, Molas2019, LiuGate, LiuValley, Liu2020, He2020, Robert2021, Robert2021PRL}. 
The detailed attribution of all emission lines is described in Section S1 of the Supplementary Information (SI), while here we focus on the two emission lines of interest that drive the functionality of our HS.
These two resonances are denoted X$^\textrm{B}$ and X$^\textrm{D}$ are associated with recombination processes of the neutral bright and dark excitons composed of carriers from the K$^\pm$ points of the Brillouin zone (BZ)~\cite{Molas2017, Koperski2017}, respectively.

\begin{figure}[t!]
		\subfloat{}%
		\centering
		\includegraphics[width=1.0\linewidth]{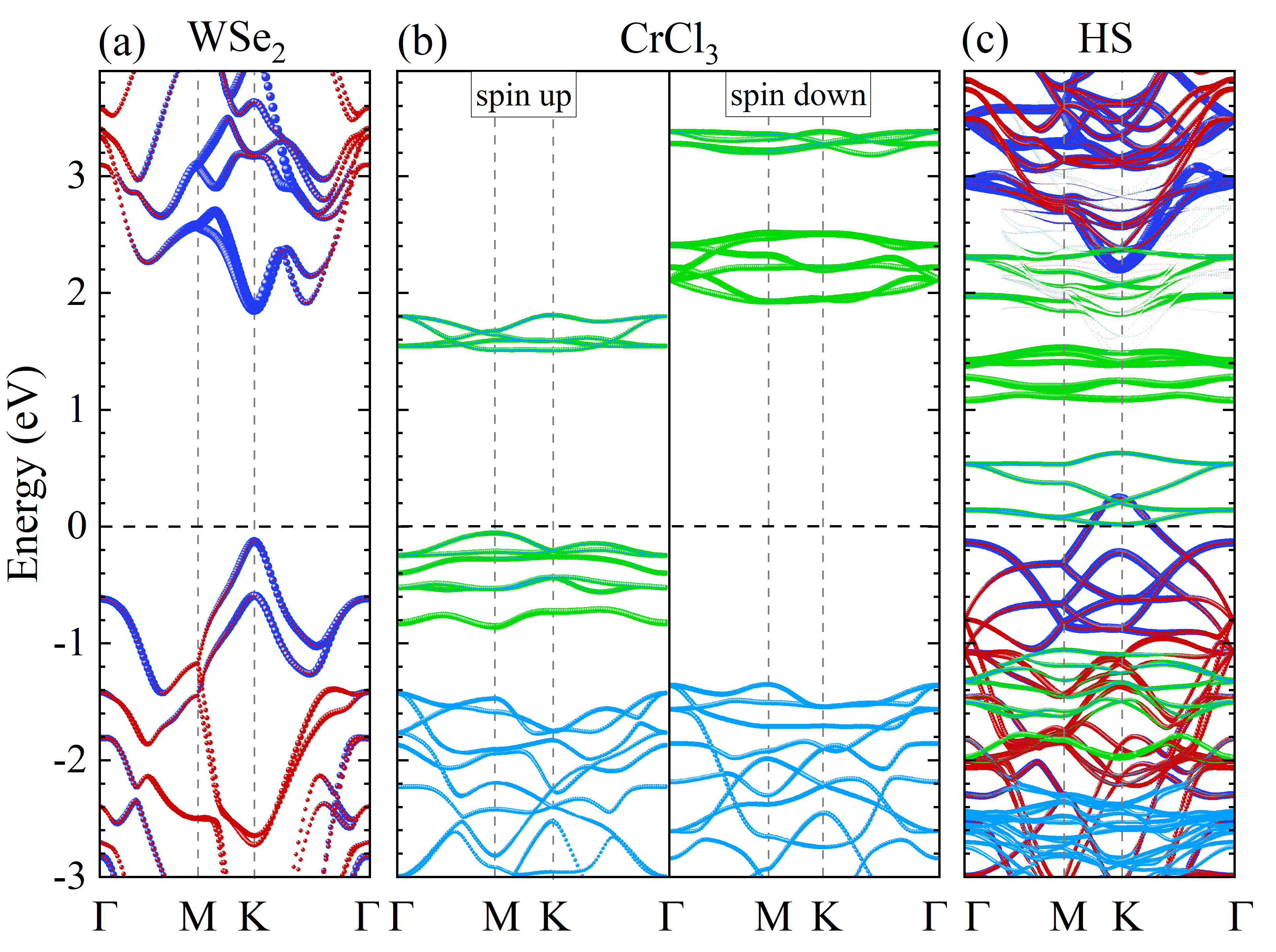}
    	\caption{The element-projected band structure of (a) WSe$_2$ monolayer, (b) CrCl$_3$ bulk, and (c) WSe$_2$/CrCl$_3$ heterostructure. The atomic projection is color-coded as follows: W~(blue), Se~(red), Cr~(light-blue), and Cl~(light-green). The band structure of WSe$_2$ ML was rigidly adjusted to fit a band gap value of 1.87 eV, which corresponds to the experimentally obtained value of the single-particle band gap based on the series of excited excitonic states. In the heterostructure (HS) case, the lattice of WSe$_2$ monolayer was stretched to match the Brillouin  zone high symmetry points between both materials for ease of comparison.}
		\label{fig:band}
\end{figure}

For the hBN/WSe$_2$/CrCl$_3$/hBN HS2, which PL spectra are analyzed in Section S2 of the SI, we observed a quenching of the total spectrally-integrated PL intensity from the WSe$_2$ ML by a factor of about 18.
The quenching is so robust that the WSe$_2$ PL signal from the majority of the hBN/WSe$_2$/CrCl$_3$/hBN HS area is negligible. 
To understand the mechanism behind the quenching of the PL intensity, we investigated the relative band alignment between bulk CrCl$_3$ and WSe$_2$ ML through density functional theory (DFT) at the level of single-particle PBE functional (see the Methods section for the computational details). 
The atom-projected band structures for the WSe$_2$ ML, CrCl$_3$, and CrCl$_3$/WSe$_2$ HS are presented in Fig.~\ref{fig:band}. 
The direct band gap at the K point of the BZ for WSe$_2$ ML was reproduced in our calculations with the electronic states around the band edges composed mainly of the d-type orbitals of tungsten. 
The spin-orbit splitting of the valence band maximum was found from first principles to yield 469 meV consistently with previous calculations using the same level of functional~\cite{Le2015}. 
Commonly, DFT methods tend to underestimate the value of the band gap. 
In our calculations, we rigidly adjusted the band gap to a value of 1.87~eV in agreement with experimentally found single particle band gap based on a ladder of excited excitonic states~\cite{Stier2017, Molas2019Energy}. 

In the case of bulk CrCl$_3$, the electronic band edges are composed predominantly of the d-type orbitals of the chromium atoms. 
The nearest neighbor Cr-Cr spin exchange results in significant splitting of the opposite spin subbands, arising from the ferromagnetic nature of CrCl$_3$. The magnetic moment per Cr atom, responsible for the emergence of net magnetization, was calculated to be equal to 3.0 $\mu_\textrm{B}$ ($\mu_\textrm{B}$ is the Bohr magneton). 

The formation of CrCl$_3$/WSe$_2$ HS and the inspection of its electronic properties through DFT methods is non-trivial as the materials exhibit a large difference of the lattice constant. 
Therefore, in terms of the absolute values of the momentum, the K points in both materials are strongly displaced. 
However, since the CrCl$_3$ bands are flat given the strong contribution of the d-type orbitals of the heavy transition metal atoms, the momentum mismatch does not lead to significant energy modifications in plausible charge transfer processes. 
Therefore, we have stretched the WSe$_2$ lattice to achieve comparable unit cells for ease of comparison. 
Under such assumptions, we obtained a band structure of WSe$_2$ ML and bulk CrCl$_3$ HS, see Fig~\ref{fig:band}(c). 
The relative position of the band edges is determined by the difference in the work function characterizing both materials individually. 
We have calculated the work function to be 6.17 eV for bulk CrCl$_3$ and 5.10 eV for WSe$_2$ ML. 
From this result, we expect a tendency to transfer electrons from the WSe$_2$ ML to the bulk CrCl$_3$. 
Indeed, from the band structure of the CrCl$_3$/WSe$_2$ HS, we can observe that the spin-up conduction band edge in CrCl$_3$ is located within the top of the ML WSe$_2$ valence band. 
Such an arrangement of the subbands leads to depopulation of the valence band of WSe$_2$ in favor of doping the conduction band of CrCl$_3$, preventing intra-WSe$_2$ optical excitations. 
The photoexcited electrons-hole pairs involving higher energy subbands are also likely to undergo recombination/relaxation through channels involving states in CrCl$_3$ at a level of excited and ground exciton states. 
Therefore, we associate the observed PL quenching in the ML WSe$_2$ with the charge transfer from WSe$_2$ to CrCl$_3$.

As the vdW interfaces are rarely perfect, we can identify individual scarce locations at several HSs, when the quenching is notably suppressed. 
At these particular locations, the character of the WSe$_2$ ML spectrum is drastically altered, as compared to the typical PL measured on the HS (see Section S2 of the SI). 
The PL spectrum becomes dominated by a single resonance at an energy of 1.68~eV, akin to the PL spectra observed at a large in-plane magnetic field that activate dark excitons via mixing of spin states between K$^+$ and K$^-$ valleys. 
We attribute this type of spectra to the features of interfacial van der Waals topography, such as bubbles or wrinkles, which spatially and electronically decouple the adjacent layers suppressing the efficiency of the charge-transfer mechanism. 
We attribute the narrow resonance to the neutral dark excitons state ($\textrm{X}^\textrm{D}$) activated by the in-plane component of the proximity field from the planar ferromagnet, initially based on the emission energy.

\begin{figure}[t!]
		\subfloat{}%
		\centering
		\includegraphics[width=1 \linewidth]{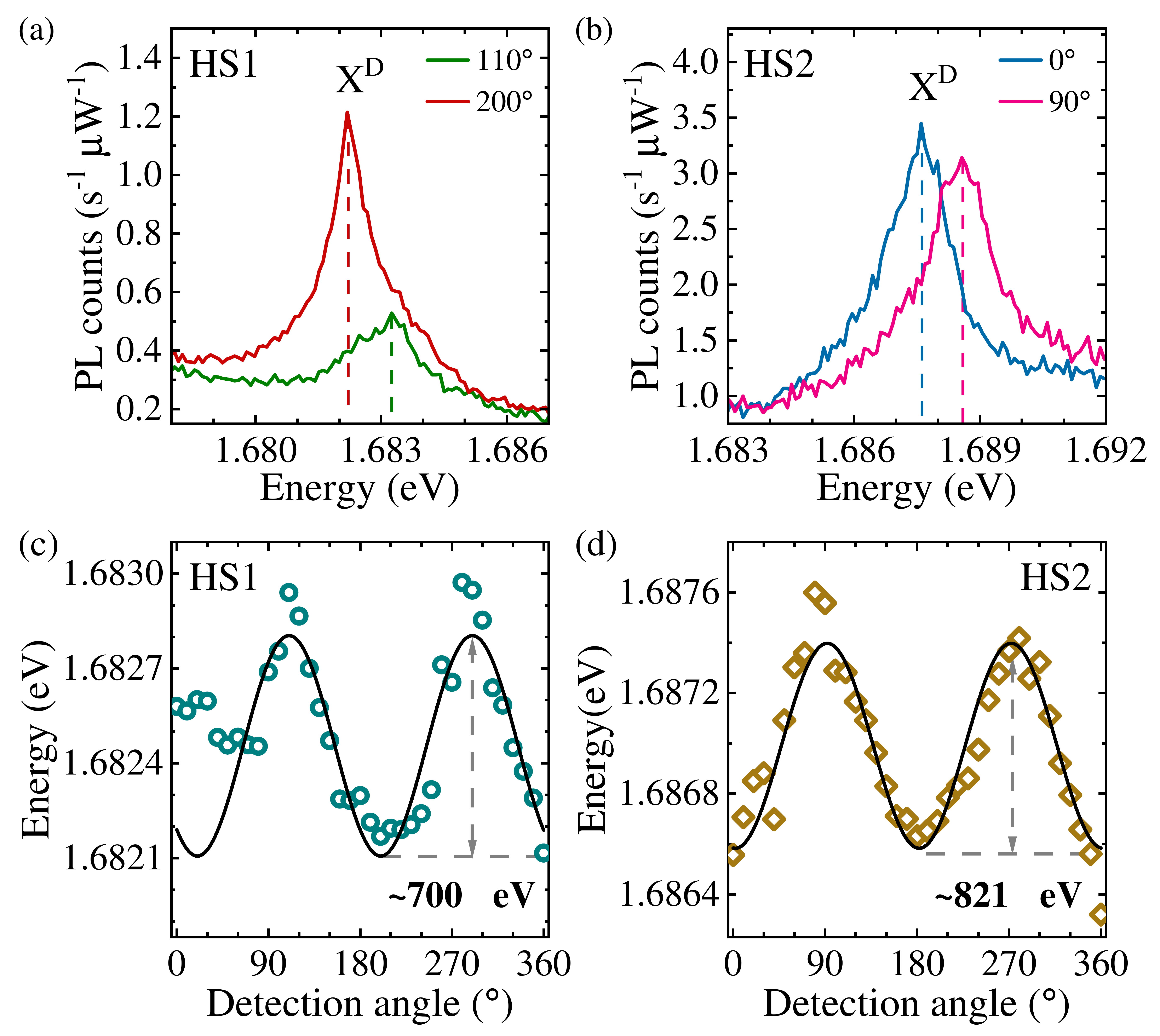}
    	\caption{Low temperature ($\textrm{T}~=~5~\textrm{K}$) photoluminescence spectra of the dark exciton X$^\textrm{D}$ resonance recorded for two orthogonal linear polarizations measured for two hBN/WSe$_2$/CrCl$_3$/hBN heterostuctures (HS): (a) HS1 and (b) HS2. 
        The corresponding linear polarization dependence of the dark exciton energy is presented for (c) HS1 and (d) HS2 samples. The solid black curves demonstrate the results of the least square fitting method using Eq.~\ref{eq:energy}.}
		\label{fig:pol}
\end{figure}

To verify our attribution of the X$^\textrm{D}$ line, we performed polarization-resolved measurements of the two similar X$^\textrm{D}$ lines identified for two HS, labeled HS1 and HS2. 
Figs.~\ref{fig:pol}(a) and (b) show the PL spectra of the X$^\textrm{D}$ lines recorded for two orthogonal linear polarizations measured in the HS1 and HS2 samples.
We found that the $\textrm{X}^\textrm{D}$ line is characterized by two linearly polarized components. 
The detailed evolution of the X$^\textrm{D}$ emission energy with the detection angle, $E(\theta)$, is shown in Figs.~\ref{fig:pol}(c) and (d). 
The line intensity was obtained based on the Lorentzian function fitting as an integral of the resonance. 
The angle dependence of the X$^\textrm{D}$ energy can be analyzed using a formula that reads
\begin{equation} 
\label{eq:energy} 
E(\theta) = E_0 + \Delta \cos^2{(\theta-\phi)},
\end{equation}
where $E_0$ and $\phi$ are the fitting parameters that describe energy and phase, while $\Delta$ represents the energy separation between two linearly polarized components of X$^\textrm{D}$. 
The fitted curves are shown in Figs.~\ref{fig:pol}(c) and (d). 
The values of the $\Delta$ parameters were $700~\pm~1~\upmu$eV for HS1 and $820~\pm~1~\upmu$eV for HS2. 
The polarization dependence and two-component structure is qualitatively consistent with previous analysis of the polarization properties of $\textrm{X}^\textrm{D}$ in the presence of in-plane magnetic fields, however, quantitatively the value of the splitting is enhanced in our structures.

\begin{figure}[h!t]
		\subfloat{}%
		\centering
		\includegraphics[width=1 \linewidth]{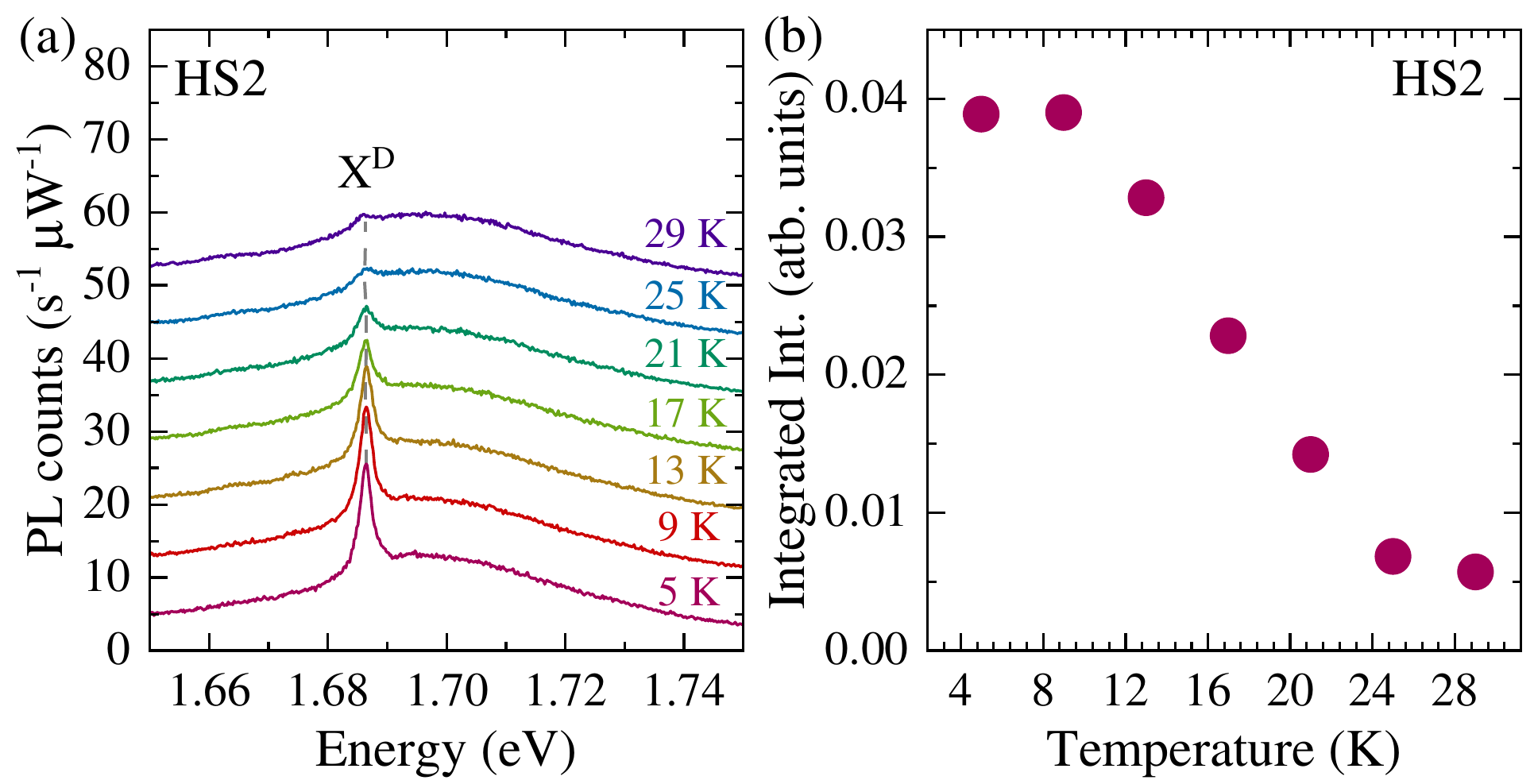}
    	\caption{(a) The temperature dependent photoluminescence spectra of the hBN/WSe$_2$/CrCl$_3$/hBN heterostructure (HS2) in range of temperatures $5~\textrm{K}~-~29~\textrm{K}$. Spectra were shifted vertically for clarity. (b) The integrated intensity of the dark exciton X$^\textrm{D}$ resonance is shown as a function of temperature.}
		\label{fig:temp}
\end{figure}

To conclude our observations, we verified the origin of the dark exciton brightening via inspection of the temperature dependence of its intensity, as demonstrated in Figs.~\ref{fig:temp}(a) and (b).
The $\textrm{X}^\textrm{D}$ feature disappears at a temperature of about $30~\textrm{K}$ which corresponds to the Curie temperature $\textrm{T}_C=27~\textrm{K}$ for bulk CrCl$_3$, which corresponds to the in-plane ferromagnetic coupling between the magnetic moments of the Cr atoms ~\cite{McGuire2017}.


\section{Discussion \label{sec:Discussion}}
The intravalley K exciton composed of carriers occupying band edges, split by spin-orbit interactions, is characterized by the opposite spin of electrons and holes in WSe$_2$ ML ~\cite{Robert2017, Molas2019, Liu2020, He2020}. 
Therefore, in principle, the oscillator strength for its recombination process should vanish. 
However, the spin-spin exchange interaction gives rise to a fine structure splitting into two types of states, which are referred to as gray and dark excitons~\cite{Slobodeniuk2016, Robert2017}. 
These two states are qualitatively different. 
The gray exciton has an optically active recombination channel with photons emitted within the plane of the ML~\cite{Wang2017}. 
This exciton can be observed in standard confocal out-of-plane configurations when using objectives with a high numerical aperture, in the absence of a magnetic field~\cite{Robert2017, Molas2019, Zinkiewicz2022}. 
The dark exciton state is truly optically forbidden, and its activation requires a magnetic field, which leads to mixing of intervalley states with opposite spins (out-of-plane configuration of the applied magnetic field)~\cite{Robert2017} or admixture of bright states into dark ones (in-plane arrangement of the external magnetic field)~\cite{Molas2017}. 
Consequently, the oscillator strength is inherited from the gray or the bright exciton, respectively. 
The exchange interactions give rise to a fine structure energy splitting between the gray and dark excitons, which are active in mutually orthogonal linear polarizations~\cite{Slobodeniuk2016, Wang2017, Molas2019, Zinkiewicz2020}. 
Our zero-field polarization-resolved inspection of the narrow resonance in the CrCl$_3$/WSe$_2$ HS indicates that we observe both gray and dark exciton states. 
The temperature dependence performed for the linear polarization, for which the dark exciton is active, demonstrates that the dark exciton arises due to the proximity magnetic field from the CrCl$_3$ ferromagnet.
Note that the polarization- and temperature evolutions of the gray exciton emission in the WSe$_2$ ML are described in Sections 4 and 5 of the SI, respectively. 
The demonstration of the origin of the modification of the PL spectra of WSe$_2$ ML allows us to inspect the functionalities of the HS in the domain of magnetic field sensing. 
Firstly, the presence of the dark exciton requires strong in-plane component of the magnetic field, typically above $1~\textrm{T}$. As the intensity of the dark exciton line increases with the magnetic field due to enhanced mixing with the bright state, the ratio of the dark exciton to the total emission from the WSe$_2$ ML can act as an estimation of the field component. 
Consequently, we estimated the in-plane component of the field to yield $2.2~\textrm{T}~\pm~0.6~\textrm{T}$ for HS1 and $1.3~\textrm{T}~\pm~0.7~\textrm{T}$ for HS2, based on the comparison with the magnetic field dependence of the PL spectrum of hBN encapsulated WSe$_2$ ML.

Our data also indicate the presence of the out-of-plane component of the magnetic field within the HS. 
The gray and dark excitons constitute two Zeeman-split branches with a zero-field anticrossing originating from the exchange coupling~\cite{Robert2017, Molas2019}. 
Therefore, the evolution of the energy of both resonances in an out-of-plane magnetic field can be described by the formula:
\begin{equation} 
\label{field_value} 
E_{G/D} = E_0 \pm \frac{1}{2} \sqrt{\Delta^2 + (g \mu_B B_\perp)^2 },
\end{equation}
where $E_G$ and $E_D$ are the energy of the gray and dark excitons corresponding to the $+$ and $-$ signs, respectively. 
$\Delta$ is the zero field splitting between the gray and dark excitons, $g$ denotes the $g$-factor, and $B_ \perp$ is the value of the out-of-plane magnetic field.

Taking values of $\Delta$=650~$\upmu$eV and $g$=9.6 from Ref.~\cite{Molas2019} and reformulating Eq.~\ref{field_value}, we obtained $B_\perp$=$(0.5\pm0.2)$~T for HS1 and $B_\perp$=$(0.9\pm0.1)$~T for HS2.

\begin{figure}[h!t]
		\subfloat{}%
		\centering
		\includegraphics[width=1\linewidth]{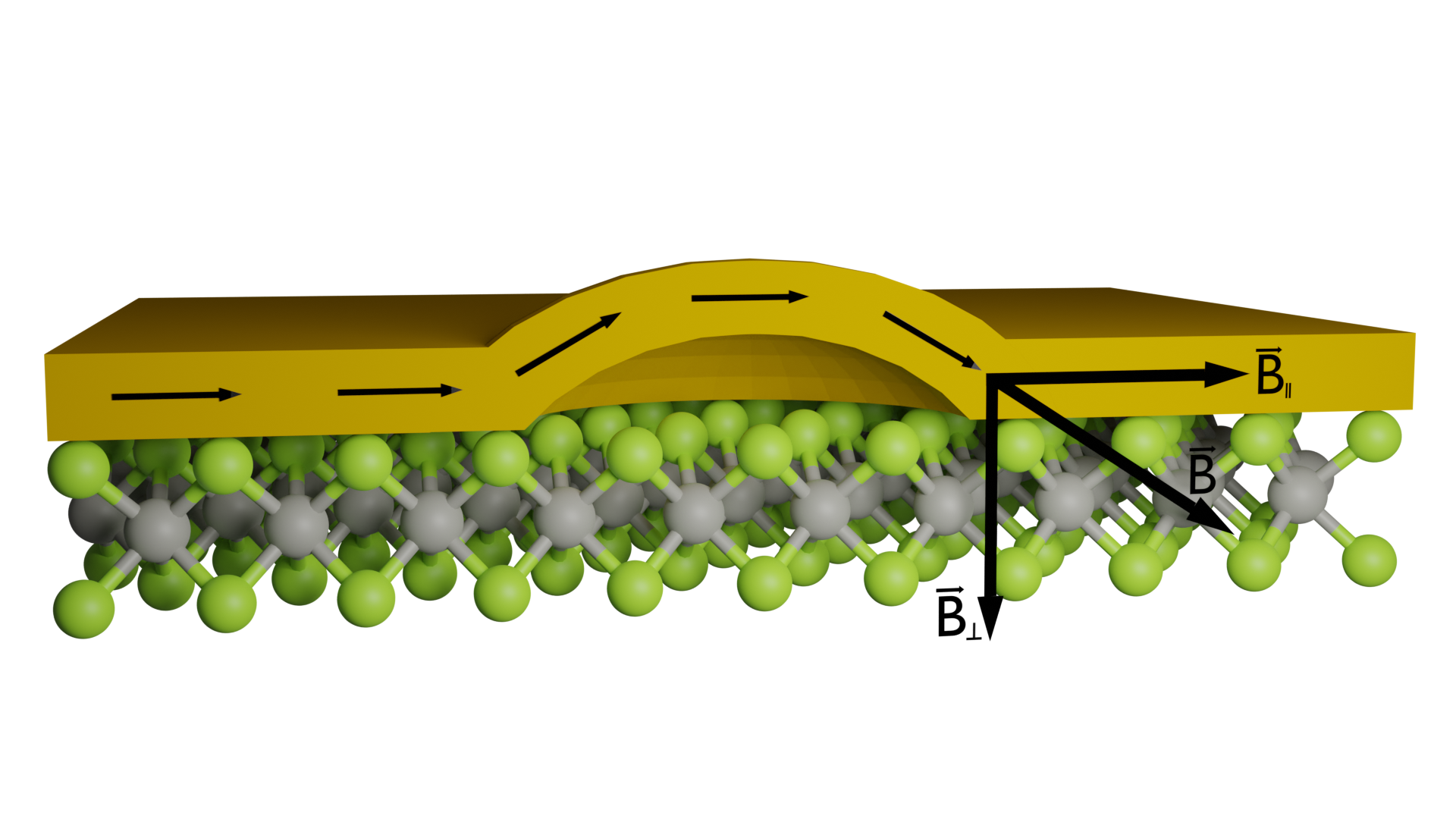}
        \caption{Simplified side view scheme of the CrCl$_3$/WSe$_2$ heterostructure. 
        The CrCl$_3$ flake is denoted by the yellow layer, and the gray and green balls depict to the W and Se atoms, respectively.
        The $\vec{B}$ represents the orientation of the effective proximity magnetic field propagating from the CrCl$_3$ layer to the WSe$_2$ ML.
        $\vec{B_\parallel}$ and $\vec B_\perp$ correspond the decomposition of the $\vec{B}$ field into the in-plane and out-of-plane components.}
		\label{fig:magnetic}
\end{figure}

Our methodology of determining the in-plane and out-of-plane components of the proximity magnetic fields enables simultaneous sensing in two dimensions, obtaining the planar and perpendicular field components with respect to the WSe$_2$ ML plane. 
We found canted orientations of the proximity field that yield $10^{\circ} \pm 2^{\circ}$ for HS1 and $28^{\circ} \pm 1^{\circ}$ for HS2. 
Based on the analysis of the PL intensity quenching mechanism, the localized nature of the brightening spots for the dark excitons and the canted orientations of the proximity field, we expect that our observations are enabled by topography features at the interface, as schematically illustrated in Fig.~\ref{fig:magnetic}.

\section{Summary \label{sec:Summary}}
In summary, we have demonstrated a ferromagnetic proximity effect within hBN/WSe$_2$/CrCl$_3$/hBN heterostructures, which led to the development of multidimensional detection of magnetic fields via inspection of a dark exciton state. 
A photoluminescence quenching mechanism via intermaterial charge transfer enabled us to isolate individual locations within the heterostructure, where the spectra are dominated by a brightened dark exciton state under zero-external-field conditions. 
This finding unlocks the potential of utilizing the long-lived dark states for synergistic optoelectronic and sensing applications via design of quantum material heterostuctures.

\section*{Methods \label{sec:methods}}
\subsection{Sample and experimental setup}
The CrCl$_3$ and WSe$_2$ crystals, used for the preparation of the investigated samples, were purchased from HQ graphene. 
Thin CrCl$_3$ and monolayer WSe$_2$ flakes were exfoliated directly on 285 nm SiO$_2$/Si substrates in an inert gas glovebox (O$_2$<1~ppm, H$_2$O<1~ppm). 
Then, we used a poly (bisphenol A carbonate)/polydimethylsiloxane stamp on a glass slide to pick up $\sim$50~nm hBN, $\sim$10~nm CrCl$_3$, monolayer WSe$_2$, ~50 nm hBN at 80$^{\circ}$C with the assistance of the transfer stage in the glove box. The hBN encapsulation limited the degradation of the CrCl$_3$ layers\cite{CrX3_degradation}. Finally, the stack was released on a fresh silicon substrate without additional layers of oxides. The thicknesses of the flakes were first identified by optical contrast and then more precisely measured with an atomic force microscope.

PL measurements were performed with the samples placed on a cold finger of a continuous-flow cryostat. 
Measurements were taken at a low temperature of $T$=5~K and as a function of temperature from $T$=5~K to 60 K using a $\lambda$ = 515 nm (2.41~eV) diode laser. 
The excitation light was focused by means of a 50$\times$ long-working-distance objective with a 0.55 numerical aperture (NA) producing a spot of about 1 $\mu$m diameter. 
The signal was collected via the same objective (back-scattering geometry), sent through a 0.75 m monochromator, and then detected by using a liquid nitrogen-cooled charge-coupled device (CCD). 
The polarization-resolved PL spectra were analyzed by a motorized half-wave plate and a fixed linear polarizer mounted in the detection path.

\subsection{DFT calculations}
Our calculations are based on DFT using the PBE functional as implemented in the Vienna Ab Initio Simulation Package (VASP) \cite{Pedrew1996, Kresse1996, Kresse1999}.
The interaction between the valence electrons and ionic cores is described within the projector augmented (PAW) approach with a plane-wave energy cutoff of 500 eV \cite{Blochl1994}.
Spin polarization was included for all the calculations. 
The lattice constant for CrCl$_3$ and WSe$_2$ is 6.06~\AA and 3.17~\AA, respectively. 
Because of the lattice mismatch between the crystal, a large moir\'e period should be expected for the interfaces of the two materials. 
The simulation of such full-scale moire heterostructure generally exceeds the ability of DFT calculation due to the large supercell and the computational consumption. 
Herein, we constructed the CrCl$_3$/WSe$_2$ heterostructure by combining a (1x1x1) CrCl$_3$ and a (2x2x1) WSe$_2$. 
The BZ was sampled using a (7x7x1) Monkhorst-Pack grid. 
A 20~\AA vacuum space was used to avoid interaction between neighboring layers. 
In the structural energy minimization, the atomic coordinates are allowed to relax until the forces on all the atoms are less than 0.01 eV/~\AA. 
The energy tolerance is 106 eV.

\section{Acknowledgments }
The authors thank Tomasz Badalski for helping with preparation of the figure presenting the side view scheme of the CrCl$_3$/WSe$_2$ heterostructure. 
The work has been supported by the National Science Centre, Poland (grant no. 2020/37/B/ST3/02311), the Ministry of Education (Singapore) through the Research Centre of Excellence program (grant EDUN C-33-18-279-V12, I-FIM) and under its Academic Research Fund Tier 2 (MOE-T2EP50122-0012), and the Air Force Office of Scientific Research and the Office of Naval Research Global under award number FA8655-21-1-7026. 
P.H. thanks the support of the National Key Research and Development Program (No. 2021YFB3802400) and the National Natural Science Foundation (No. 52161037) of China.
K.W. and T.T. acknowledge support from the JSPS KAKENHI (Grant Numbers 20H00354 and 23H02052) and World Premier International Research Center Initiative (WPI), MEXT, Japan. 
The computational work are performed on computational resources at the National Supercomputing Center of Singapore (NSCC) and the NUS HPC.

\bibliographystyle{apsrev4-2}
\bibliography{biblio}

\newpage
\onecolumngrid
\setcounter{figure}{0}
\setcounter{section}{0}
\renewcommand{\thefigure}{S\arabic{figure}}
\renewcommand{\thesection}{S\Roman{section}}
	\begin{center}
	{\large{{\bf  \textsc{Supplementary Information:}} \\ Multidimensional sensing of proximity magnetic fields via intrinsic activation of dark excitons in WSe$_2$/CrCl$_3$ heterostructure}}
\end{center}

\section{Low-temperature PL spectrum of WSe$_2$ monolayer \label{sec:S1}}

\begin{figure}[h]
		\subfloat{}%
		\centering
		\includegraphics[height=2.3in]{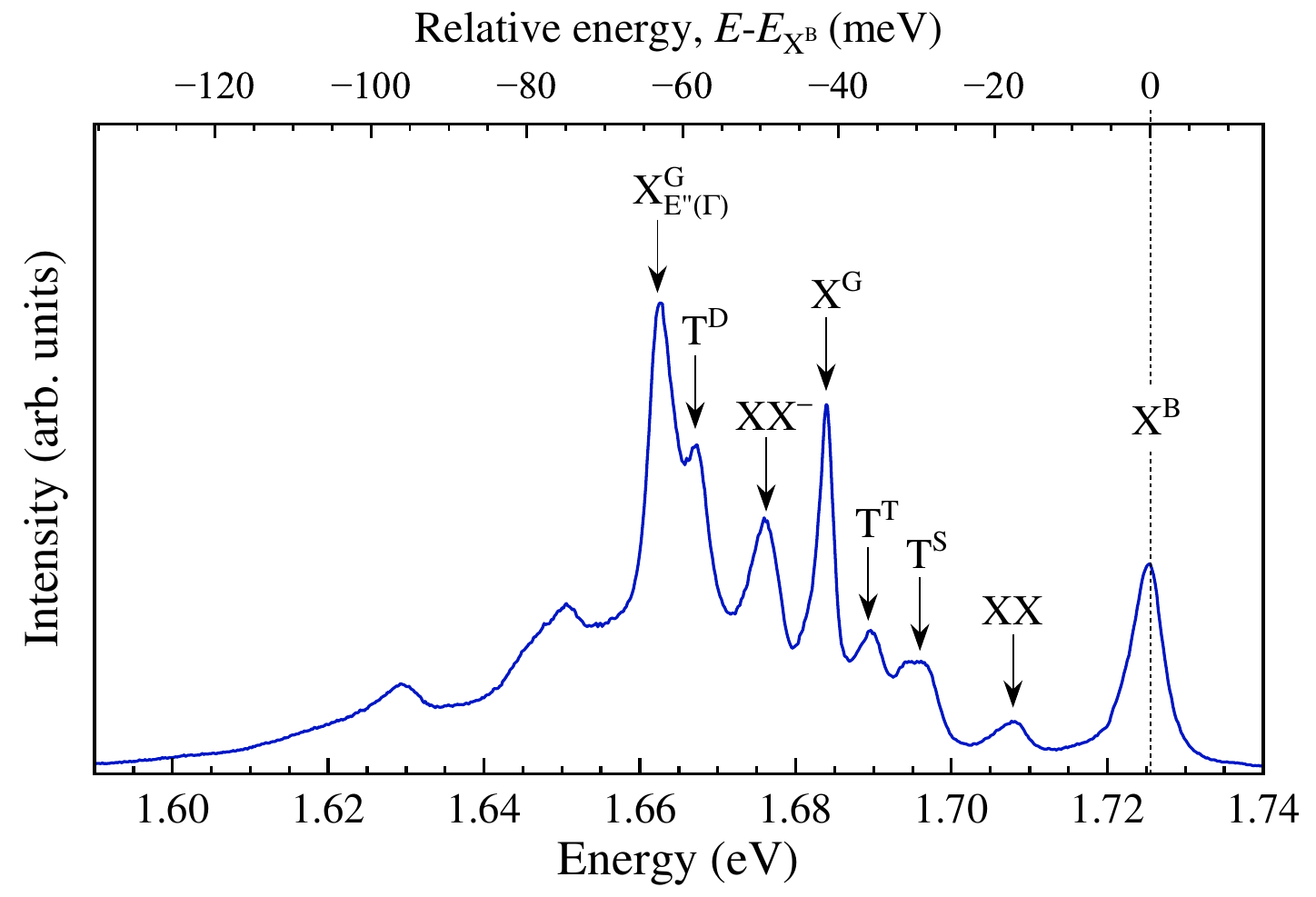}
    	\caption{Low-temperature ($T$=5~K) PL spectrum of the WSe$_2$ ML encapsulated in hBN flakes measured using excitation energy 2.41~eV and power of 20~$\mu$eV. 
        Lines assignments is a follows: X$^\textrm{B}$ -- neutral exciton; 
        XX -- neutral biexciton; 
        T$^\textrm{S}$ and T$^\textrm{T}$ -- singlet (intravalley) and triplet (intervalley) negatively charged excitons, respectively; 
        X$^\textrm{G}$ -- gray exciton, 
        XX$^-$ -- negatively charged biexciton; 
        T$^\textrm{D}$ -- negatively charged dark exciton (dark trion); 
        E$^\textrm{G}_{\textrm{E}"(\Gamma)}$ -- phonon replica of gray exciton.}
		\label{fig_WSe2}
\end{figure}

Fig.~\ref{fig_WSe2} shows the low-temperature ($T$=5~K) PL spectrum measured on a WSe$_2$ ML encapsulated in hexagonal BN (hBN) ﬂakes.
The spectrum displays several emission lines with a characteristic pattern similar to that previously reported in several works on WSe$_2$ MLs embedded in between hBN ﬂakes~\cite{Courtade2017, Li2018, Chen2018, Barbone2018, Paur2019, Liu2019, Li2019, Li2019replica, Li2019momentum, Molas2019, LiuGate, LiuValley, Liu2020, He2020, Robert2021, Robert2021PRL}. 
According to these reports, the assignment of the observed emission lines is as follows: 
X$^\textrm{B}$ -- neutral exciton; 
XX -- neutral biexciton; 
T$^\textrm{S}$ and T$^\textrm{T}$ -- singlet (intravalley) and triplet (intervalley) negatively charged excitons, respectively; 
X$^\textrm{G}$ -- gray exciton, XX$^-$ -- negatively charged biexciton; 
T$^\textrm{D}$ -- negatively charged dark exciton (dark trion); 
E$^\textrm{G}_{\textrm{E}"(\Gamma)}$ -- phonon replica of gray exciton due to the emission of the E" phonon from the $\Gamma$ point of the Brillouin zone.

\newpage
\section{Photoluminescence spectra of heterostructure HS2 \label{sec:S4}}

\begin{figure*}[h]
		\subfloat{}%
		\centering
		\includegraphics[width=1 \linewidth]{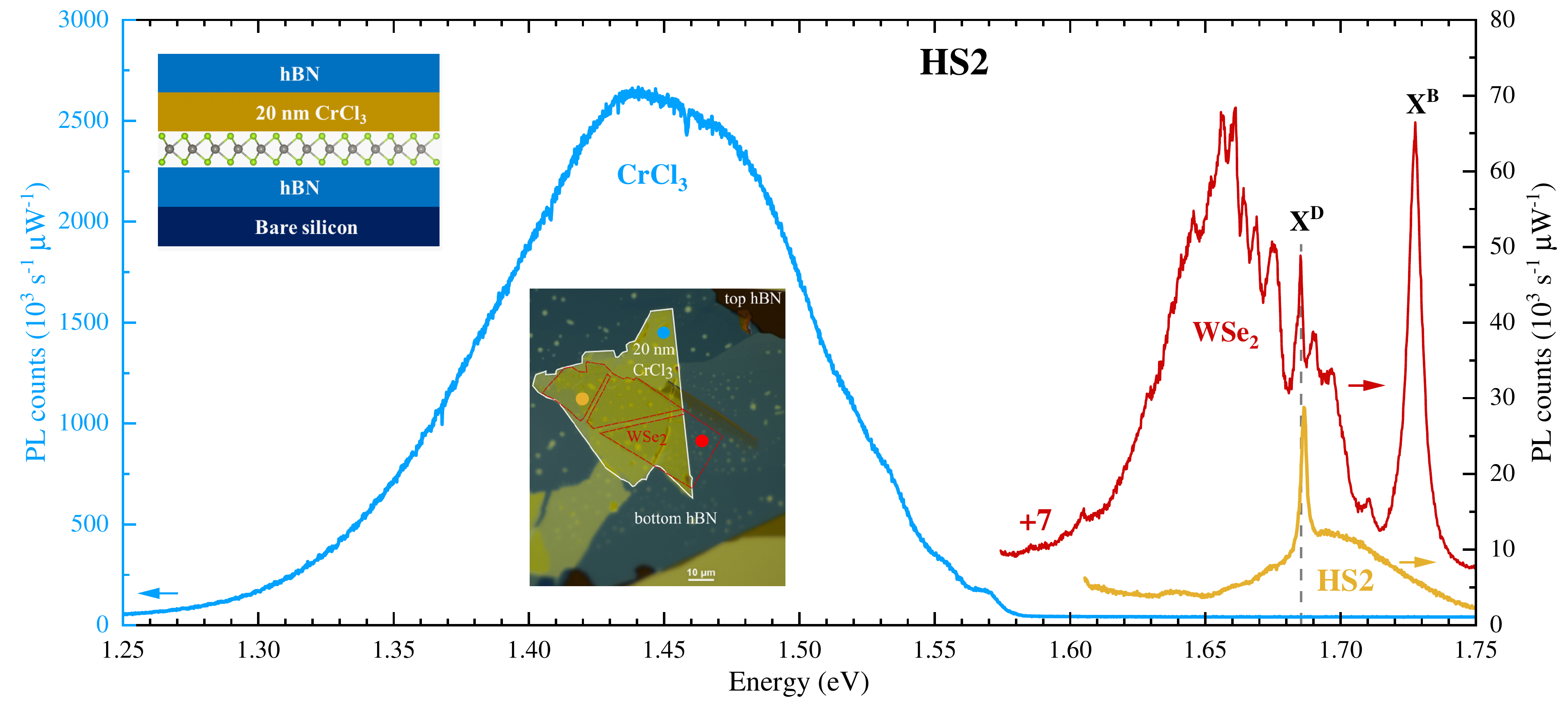}
    	\caption{PL spectra of three structures: hBN/CrCl$_3$/hBN with a 20 nm thick CrCl$_3$ film (blue), a hBN/WSe$_2$/hBN with WSe$_2$ monolayer (red) and heterostructure hBN/WSe$_2$/CrCl$_3$/hBN (yellow) measured on the sample referred to as HS2 at low temperature ($\textrm{T}=5~\textrm{K}$), using excitation energy 2.41~eV and laser power of 15~$\upmu$W. 
     The WSe$_2$ spectrum is vertically shifted for clarity.
     The inset in the top-left corner shows the schematic side view of the sample with the indication of consecutive layers. 
     The inset in the central part of the figure demonstrates the optical image of the heterostucture. 
     The outlines for individual flakes are added as a guide to the eye. 
     Colored dots in the optical image indicate spatial positions corresponding to the PL spectra of different material configurations.}
		\label{fig_HS2}
\end{figure*}

\begin{figure*}[h]
		\subfloat{}%
		\centering
		\includegraphics[width=0.65 \linewidth]{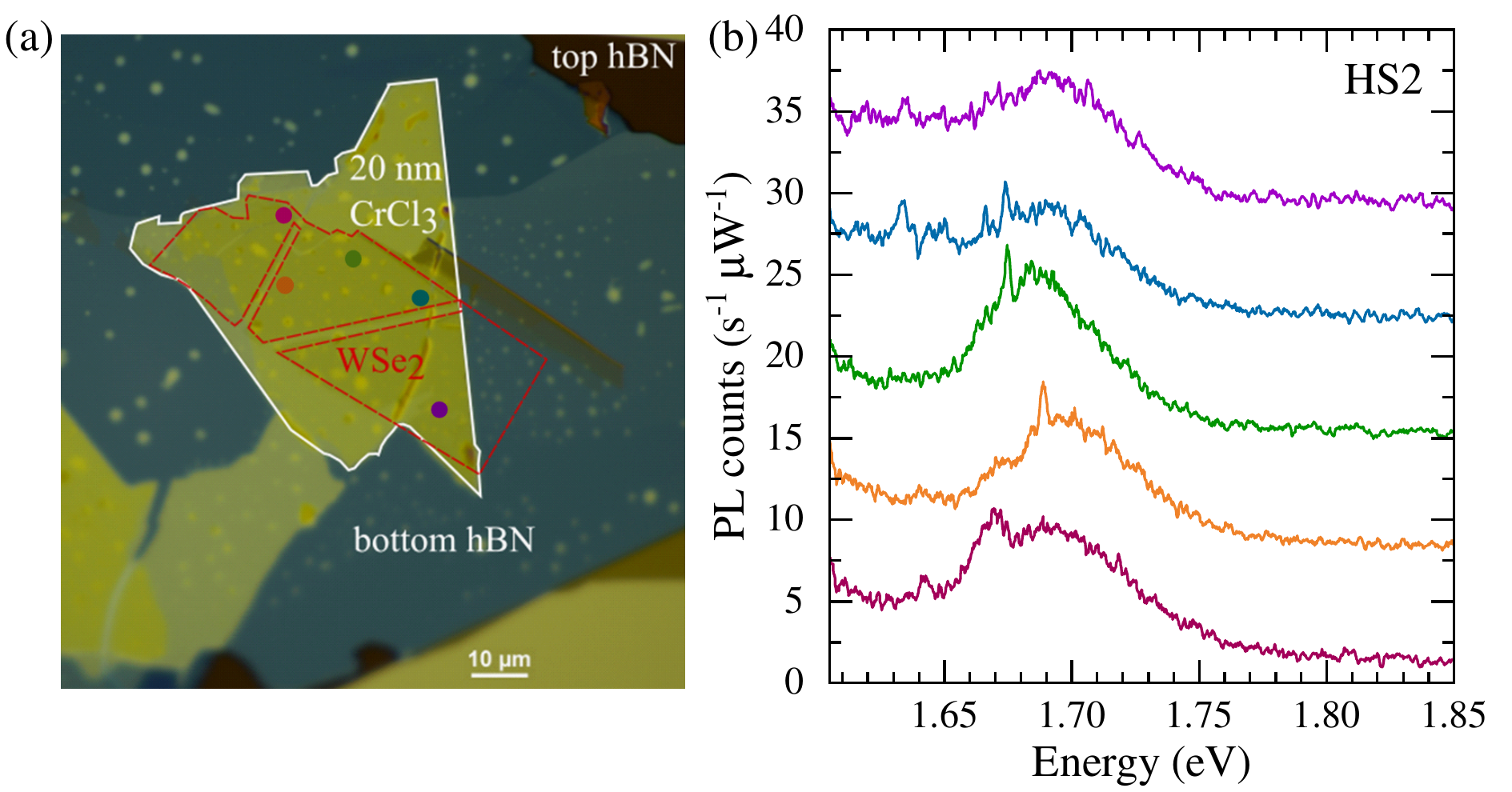}
    	\caption{(a) An optical image of HS2 with selected measured areas marked by colored dots that correspond to the photoluminescense spectra (b) of structure: hBN/WSe$_2$/CrCl$_3$/hBN measured at low temperature ($\textrm{T}=5~\textrm{K}$), using excitation energy 2.41~eV and laser power of 15~$\upmu$W. 
        The spectra are vertically shifted for clarity.}
		\label{fig:S5}
\end{figure*}

Fig. \ref{fig_HS2} shows the low-temperature ($T$=5~K) PL spectra measured on the sample HS2.
It consists of the PL spectra of the 20-nm-thick CrCl$_3$ flake, the encapsulated WSe$_2$ ML, and the hBN/WSe$_2$/CrCl$_3$/hBN heterostructures, which are similar to those presented in the main article. 
In particular, the relative intensity of the heterostructure HS2 area to the WSe$_2$ ML is significantly enhanced compared to the presented in the main article.
Nevertheless, for the hBN/WSe$_2$/CrCl$_3$/hBN HS, we observed a significant quenching of the total spectrally-integrated PL intensity from the WSe$_2$ ML by a factor of 3.7.
To visualize the quenching effect, we presents the low-temperature ($T$=5~K) PL spectra measured on five spatial locations of the hBN/WSe$_2$/CrCl$_3$/hBN heterostructures in Fig.~\ref{fig:S5}.
The shown spectra are very similar to each other and are dominated by a broad emission band located at about 1.7~eV.
This confirms the strong charge-transfer process from the WSe$_2$ ML to the CrCl$_3$ flake, as described in the main article.

\clearpage
\section{Polarization-resolved spectra of neutral and gray exciton \label{sec:S2}}

\begin{figure}[h]
		\subfloat{}%
		\centering
		\includegraphics[height=2.3in]{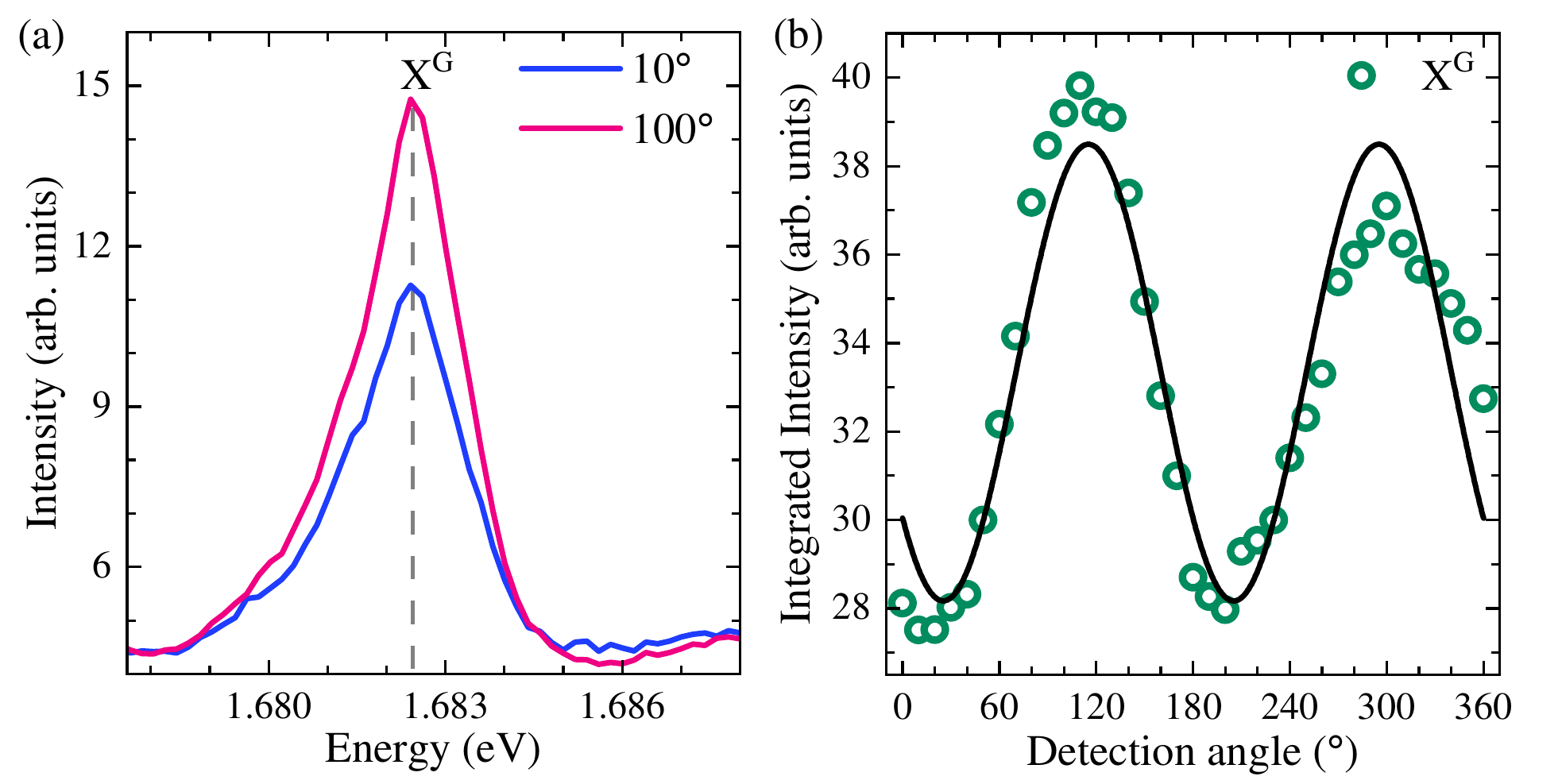}
    	\caption{(a) Low-temperature ($T$=5~K) PL spectra of the X$^\textrm{G}$ line recorded for two orthogonal linear polarizations.
        The measured angle are in reference to the fast axis of the used half-wave plate in the detection path.
        (b) Polarization dependence of the gray exciton integrated intensity (green points) as a function of detection angle.
        The solid  black line represents fit according to Eq.~\ref{eq:intensity}.}
		\label{fig_pol}
\end{figure}

To reveal the effect of magnetic proximity on the gray/dark exciton emission in the studied WSe$_2$/CrCl$_3$ heterostructures, we investigated the polarization properties of the neutral gray exciton line (X$^\textrm{G}$) apparent in the low-temperature ($T$=5~K) PL spectrum of WSe$_2$ ML encapsulated in hBN flakes.
As can be seen in Fig.~\ref{fig_pol}(a), only the X$^\textrm{G}$ intensity changes for two orthogonal detection angles, while its energy is not modified. 
It is known that the emission due to the neutral gray exciton is observed exclusively without additional in-plane and/or out-of-plane magnetic fields, whereas the emission due to the neutral dark exciton is apparent in non-zero magnetic fields~\cite{Robert2017, Molas2019}.
Despite the fact that the neutral gray exciton is characterized by the dipole momentum perpendicular to the ML plane (it should emit light only in the ML plane with its out-of-plane polarization)~\cite{Wang2017}, the X$^\textrm{G}$ emission measured on WSe$_2$ MLs was also observed using the out-of-plane configuration (the excitation/detection light is orthogonal to the ML plane) using the large numerical aperture (NA) of the microscope objective used to collect the emitted light~\cite{Robert2017, Molas2019, Liu2020, He2020}.
Moreover, for this experimental configuration, the X$^\textrm{G}$ emission is partially linearly polarized (see Ref.~\cite{Molas2019} for details).
To investigate it, we fit the integrated intensity of the gray exciton ($I$) as a function of the detection angle ($\theta$), shown in Fig.~\ref{fig_pol}(b), using formula
\begin{equation}
I(\theta)=I_0 + A \cos^2(\theta-\phi),
\label{eq:intensity}
\end{equation}
where $I_0$ is the background intensity, $A$ represents the magnitude of the intensity variation, and $\phi$ is the phase parameter.
The polarization evolution of the X$^\textrm{G}$ line can be nicely described using the formula.
This supports the idea that the gray exciton peak is composed of two linearly polarized components, which are degenerated in energy.
In conclusion, only the emission of the neutral gray exciton is observed without additional in-plane and/or out-of-plane magnetic fields, as already reported in Refs.~\cite{Robert2017, Molas2019}.

\clearpage
\section{Temperature evolution of the PL spectra of the WSe$_2$ ML \label{sec:S3}}

\begin{figure}[h]
		\subfloat{}%
		\centering
		\includegraphics[height=2.3in] {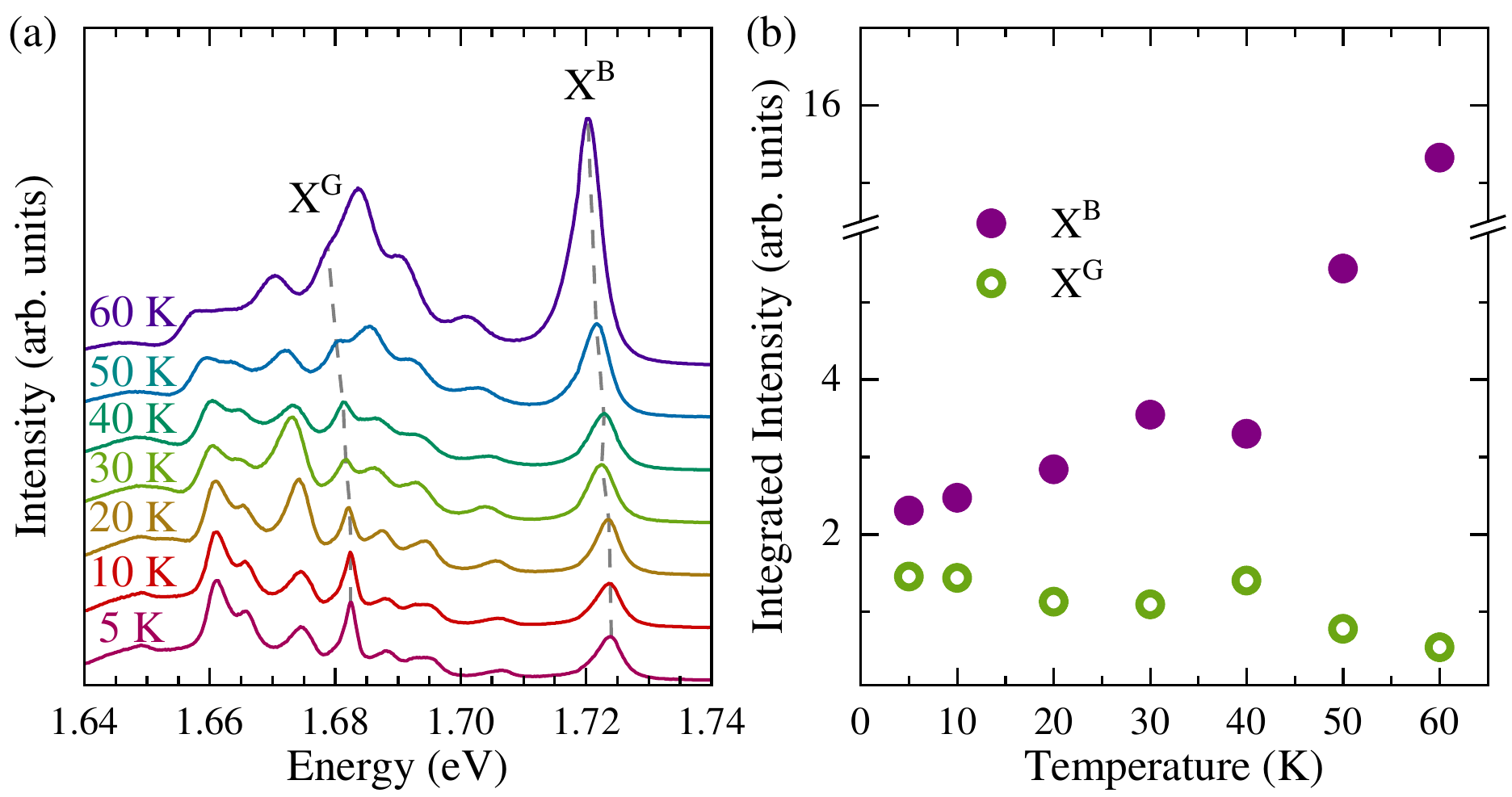}
    	\caption{(a) Temperature dependent PL spectra of WSe$_2$ encapsulated in hBN flake in the temperature range from 5 K to 60 K measured using excitation energy 2.41~eV and power of 15~$\mu$W. 
        Spectra are shifted vertically for clarity. 
        (b) Evolution of the integrated intensities of the neutral bright (X$^\textrm{B}$) and gray (X$^\textrm{G}$) excitons as function of temperature.}
		\label{fig_temp}
\end{figure}

Fig.~\ref{fig_temp}(a) shows the temperature dependent PL spectra of WSe$_2$ encapsulated in hBN flakes in the temperature range from $T$=5~K to 60~K. 
With increasing temperature, consecutive low-energy peaks in the energy range of about 1.655~eV -- 1.685~eV progressively disappear one after another from the spectrum.
At the same time, three emission lines, X$^\textrm{B}$, T$^\textrm{S}$, and T$^\textrm{T}$, gain their intensities in this temperature range. 
Note that while the gray exciton line is nicely pronounced at $T$=5~K, its intensity decrease quickly with increasing temperature, and this peak is hardly resolved at $T$=60~K.
To compare the temperature evolution of the X$^\textrm{B}$ and X$^\textrm{G}$ lines, we plot their integrated intensities as a function of temperature, see Fig.~\ref{fig_temp}.
The intensity of the X$^\textrm{B}$ line increases with temperature, as has already been reported in Ref.~\cite{Zhang2015}, while the X$^\textrm{G}$ intensity quenches above 60~K. 
The obtained results are very similar to those described in Ref.~\cite{Robert2017}, where the quenching temperature of the X$^\textrm{G}$ line was found to be of about 50~K.

\end{document}